\documentclass[twocolumn,aps,floats,floatfix,superscriptaddress]{revtex4}
\usepackage{graphicx,amssymb}
\usepackage{epsfig}
\newcommand{\beq}{\begin{equation}}
\newcommand{\eeq}{\end{equation}}

\newcommand{\bE}{{\bf E}}

\newcommand{\bZ}{{\bf Z}}

\newcommand{\bN}{{\bf N}}

\newcommand{\br}{{\bf r}}
\newcommand{\bj}{{\bf j}}

\newcommand{\ez}{{\bf e}_{z}}
\newcommand{\ep}{{\bf e}_{+}}
\newcommand{\emm}{{\bf e}_{-}}

\begin{document}
\draft
\title{Scattering of electromagnetic waves from a cone with
conformal mapping: application to scanning near-field optical microscope 
}
\author{ S. T. Chui}
\affiliation{ Bartol Research Institute and Dept. of Physics and Astronomy, 
University of Delaware, Newark, DE 19716, USA}
\author{  Mengkun Liu and Xinzhong Chen }
\affiliation{ Dept. of Physics and Astronomy, 
Stony Brook University, Stony Brook, N.Y. 11794, USA}
\author{  Zhifang Lin and Jian Zi}
\affiliation{ Dept. of Physics, 
Fudan Univ., Shanghai, China}
\begin{abstract}
%
%
%
%
%
%
%
%
%
%
%
%
%
%
%
%
%
%
%
We study the response of a conical metallic surface to an external electromagnetic (EM) field by representing the fields in basis functions containing integrable singularities at the tip of the cone. A fast analytical solution is obtained by the conformal mapping between the cone and a round disk.
We apply our calculation to the scattering- based scanning near-field optical microscope (s-SNOM) and successfully quantify the elastic light scattering from a vibrating metallic tip over a uniform sample.
We find that the field-induced charge distribution  consists of localized terms at the tip 
and the base and an extended bulk term along the body of the cone far away from the tip. 
In recent s-SNOM experiments at the visible-IR range (600nm - 1$\mu m$) the fundamental is found to be much larger than the higher harmonics whereas at THz range ($100 \mu m-3mm$) the fundamental becomes comparable to the higher harmonics. We find that the localized tip charge dominates the contribution to the higher harmonics and becomes bigger for the THz experiments, thus providing an intuitive understanding of the origin of the near-field signals. 
We demonstrate the application of our method by extracting a two-dimensional effective dielectric constant map from the s-SNOM image of a finite metallic disk, where the variation comes
from the charge density induced by the EM field. 
\end{abstract}
\maketitle
\begin{figure}[tbph]
\vspace*{0pt} \centerline{\includegraphics[angle=0,width=9cm]{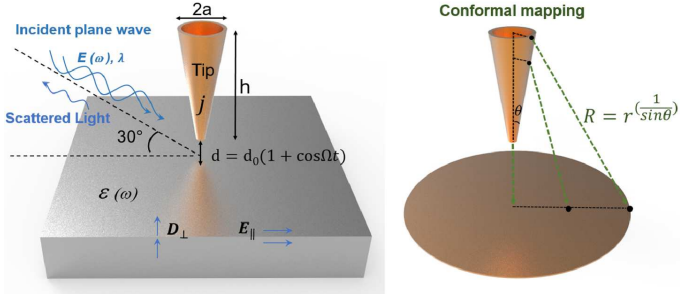}} 
\vspace*{0pt}
\caption{(a) Schematic representation of the scattering near-field setup. 
(b) Schematics of the conformal mapping of the cone shaped tip to a round disk.} 
\label{geom}
\end{figure}

Scattering-type near-field optical microscope (s-SNOM) has made great advances over the past decade by combining the well-developed atomic force microscope (AFM) techniques with a wide range of tunable and broadband light sources. 
By gathering the scattered light from an AFM tip (see Fig. \ref{geom}a), 
s-SNOM effectively probes the near-field interactions between the nanometer-sized tip apex and the sample surface, providing a spatial resolution of ~10 nm, far beyond the diffraction limit of traditional optics\cite{nov,liu}. It has shown to be extremely useful in probing phase inhomogeneities in strongly correlated electron materials\cite{1,2,3,4,5,6}, polaritonic waves in 2D materials\cite{7,8,9}, charge concentrations in semiconductor nanostructures\cite{10,11,12} and molecular nano-fingerprints in organic and soft materials\cite{13,14,15}. 
The central task of quantitatively understanding the near-field signals comes down to solving a nontrivial scattering problem, of which light of frequency $\omega$ elastically scatters from a cone-shaped AFM tip oscillating vertically on the sample surface at the mechanical resonance frequency $\Omega<<\omega$ of the cantilever.
Approaches from simple models such as the point dipole\cite{pd} and rp models\cite{rp} to state-of-the-art finite element calculations\cite{cst} have been applied to this problem.
Because of the singularity at the tip, current approaches have not been able to consistently predict the phase of the signal accurately (See Fig. 3 below). We find that the 
inverse extraction of the physical properties of the sample from the scattering data with 
current numerical approaches
does not always produce unique results. 
It is also extremely time consuming to solve the problem without simplifying approximation and
this has limited the reliable real time interpretation of the 
experimental data. 

Here we overcame the difficulty of the singularity at the apex by 
using our recently developed technique\cite{book,disk,tri} of representing the fields
in basis functions containing integrable singularities at the tip  obtained by conformal
mapping between the cone and a round disk and successfully quantify the elastic light
scattering from a vibrating metallic tip. We found the field-induced charge distribution
to consist of a bulk term along the body of the cone far away from the tip
 and localized terms at the tip
apex and the base.
The scattering at the fundamental frequency  $\omega+\Omega$ comes
from both the bulk and the localized charges. 
The near field higher harmonics at frequencies $\omega+n\Omega$ experimentally observed is dominated by contributions from the localized charge at the tip.
In recent s-SNOM experiments at the optical to mid infrared range\cite{liu} and in the 
THz range ($>1 mm$)
the fundamental is much larger than and comparable to the higher harmonics respectively. 
Our calculation is in agreement with the empirical measurements
where we found that the tip charge at THz frequencies is much larger than that at IR
frequencies.
To mimic the conical AFM tip, there have been
work with different less singular tip geometries such as the 
spherical\cite{pd}, the spheroidal\cite{17,18,19} and the pear-shaped\cite{XXX}. 
These calculations implied different induced charge distributions. Our approach provides a more
systematic way to calculate the charge distributions for different experimental conditions.

Our calculation is also orders of magnitude faster than current approaches and makes this study suitable for efficiently calculating 
the near-field signals and tip-sample interactions over a wide range of frequencies all the
way from visible to terahertz with a fast and effective theoretical treatment.
We demonstrate the back-extraction of a two-dimensional map from the s-SNOM image of a finite metallic disk which effectively quantify the charge density of a finite metallic surface induced by an EM field.
We now described our results in detail.

We are interested in the current flow on a finite film caused by an external electromagnetic (EM) wave. 
We assume that the film is thin enough that there is no current in the direction perpendicular to it. The current density ${\bf j}$  
in the presence of an external electric field $\bE_{ext}$ is governed by the equation
\beq
\rho{\bf j+ E_{em}=E}_{ext}.
\eeq
where $\rho$ is the resistivity, ${\bf E}_{em}$ is the electromagnetic field generated by the current.
We impose the boundary condition of no current flow perpendicular to the boundary of the film  with a large
boundary resistivity $\rho_s$ which we take to approach infinity\cite{book,disk,tri}. The total resistivity $\rho$ is a sum of this boundary term and a metal resistivity $\rho_0$.  

Because of the singularity at the tip, it is difficult to treat the problem.
Here we represent the currents and the fields {\bf not} in terms of
finite elements on a mesh but in terms of a complete set of orthonormal basis functions
with the integrable singularity of the system built in.
This rigorous approach was shown to be very efficient\cite{book,disk,tri}. 
 For the simple case
of an annnulus of radii $R_1$, $R_2$, the basis functions are the well known vector cylindrical
functions\cite{disk,tri} ${\bf M}_m$ ${\bf N}_m$
of angular momentum $m$ and with components $X_r$ and $X_{\Phi}$
(X=M, N) in the radial and the angular directions. 
Here we are interested in a conical film of height $h$, base radius $a$ with the cone
angle $\theta=\tan^{-1}a/h$,  thickness $t$ and a small flat tip of radius $b$.
This geometry is closest to the experimental setup among all the theoretical models.
A point on the conical surface and a point inside the annulus  are
characterized by the cylindrical  coordinates $r$ and $\phi$ and $R$ and $\Phi$
respectively. As is illustrated in Fig. \ref{geom}b, 
these can be mapped into each other via a conformal harmonic map\cite{map}
$$R=r^{1/\sin\theta},\ \Phi=\phi.$$ 
The Jacobian   of the transformation is $J =R^{2\sin\theta-2}$.
We construct the basis function for this surface from the basis function of a circular annulus
as 
\beq
{\bf cX}(r)=\alpha \left\{X_r[{\bf R}({\bf r})]{\bf e}_{1}+ X_{\Phi}{\bf e}_{2}\right\}
\label{cX}
\eeq
where $X=M,\ N$, $\alpha=1/J^{1/2}=R/r.$ 
The tangent vectors on the surface of the cone are
in the cylindrical basis $(r,\phi,z)$: ${\bf e}_1=
1/(1+h^2)^{1/2}{\bf e}_r-h/(1+h^2)^{1/2}{\bf e}_z,$ ${\bf e}_2={\bf e}_{\phi}$
With this choice, the new basis functions are orthonormal with the corresponding measure: 
$\int d^2r {\bf (cX_n)^*\cdot cY_m}=
\int d^2R J |\alpha|^2 {\bf X_n\cdot Y_m}=\delta_{X,Y}.$

The electromagnetic field $\bE_{em}$ can be represented as
${\bf E}_{em}={\bf Z^0 j}$ where the
"impedance" matrix $\textbf{Z}^0$
is just the representation of the Green's function in this basis\cite{book,disk,tri}.
Just as in our previous studies, when the basis functions are orthonormal, the off-diagonal elements of the impedance matrix is much less than the diagonal elements. Furthermore the magnitude of the impedance increases rapidly. These greatly  facilitate the convergence of the solution and provide for a much better understanding of the physics. 

In this notation, the circuit equation becomes
\beq
{\bf Zj=E}_{ext}+\bE_s
\label{cire}
\eeq
where $Z=Z^0+\rho_0$.
The boundary electric field $\bE_s$ is the product of the normal component of the current at the boundary $j_s$ and $\rho_s$, i.e. $\bE_s=\bj_s\rho_s$ . They behave like Lagrange multipliers. Their values are determined from the condition that the normal boundary currents become zero.
Physically, as the external field is applied, the current is stopped at the boundary and charges of surface density $\sigma_s$ are getting accumulated. A boundary field is generated until it reaches a value to oppose further current arriving there.


In the experiment, there is an additional sample surface of dielectric constant $\epsilon$ 
underneath the cone. 
The EM field induced charges of density $n$ on the surafce. 
In the presence of the plane at a distance $d=d_0(1-\cos\Omega t)+d_m$ from the tip, 
there
will be image charges of density $n_i(\br-2d\ez)=-\beta n(\br),$
$\beta =(\epsilon-1)/(\epsilon+1).$ Because $d$ is much less than the wavelength
the method of images is a good approximation. 
The total EM field now has additional contributions from the image charges. The effective 
impedance $\bZ$ is modified. 
We have calculated the additional image circuit parameters and included them in the circuit equations.

Because of the image charge, the tip surface field  $E_{s2}=E_{sa}+E_{sb}$ 
is now a sum of a field $E_{sa}$ from a surface charge of density 
$\sigma_s$ at the tip  and a field $E_{sb}$ from
the image charge density
$-\beta \sigma_{s}$ at a distance $-d$ below the surface.
From Gauss's law $E_{sa}=\sigma_s/2\epsilon_0,$ $E_{sb}\approx -\beta \sigma_s/\epsilon_0I(2d)$ where 
$I(x)= <E_{cz}>$ is the z Coulomb electric field from the image 
charge density on the tip averaged over the tip. 
From these equations, we get 
\beq
\sigma_s=2\epsilon_0 E_{s2}/[1-2\beta I(2d)].
\label{scd}
\eeq
We have calculated $I(x)$ numerically and tabulated it for a mesh of 4000 points in the
region $0<x/d_0<2$. 

Recent experiments have been carried out at IR 
frequency with $\lambda=10\mu m$, $h=20\mu m,$ $a=5\mu m,$ $b=10 nm,$ $d_0=60 nm,$ $t=20nm,$ $d_m=0.6 nm$ 
and at THz frequency with $\lambda$ centered at $300\mu m$, $h=70\mu m,$ $a=20\mu m,$ $b=50 nm,$, $d_0=150 nm$, $d_m=0.6nm$. In the latter case, the tip is solid; the effective thickness of the film is thus the skin depth $\xi$. For this case $t/a\approx 0.01$ whereas for the IR case, $t/a=4\times 10^{-3}.$
We have carried out calculations for these parameters.
%
The incoming field 
is at an angle of 60 degrees with respect to the surface normal. This field is nearly
uniform across the tip. The coupling is thus dominated by the m=0 mode which we focus on.  

\begin{figure}[tbph]
\vspace*{0pt} \centerline{\includegraphics[angle=0,width=8cm]{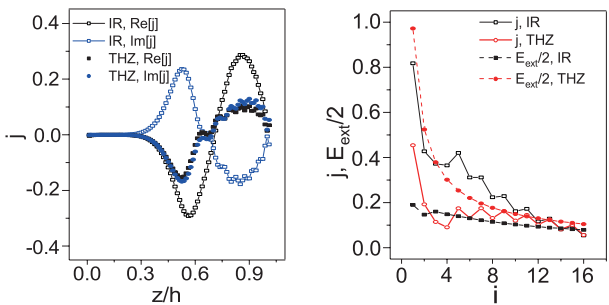}} 
\vspace*{0pt}
\caption{ The magnitude of the expansion coefficient for the bulk current density $j^c$ 
(right panel) in terms of the normal modes wave vectors index $i$
and (left panel) as a function of its normalized distance from the tip for parameters for two sets of experiments in the IR and THz regime.
(right panel) The magnitude of the expansion coefficient for the external field $E_{i,ext}$ in terms of $i$.}
\label{je}
\end{figure}

We have calculated the circuit parameters $\bZ$ for the m=0 modes with the basis functions
$cX_{m=0}(k_ir)$ for up to 16 wave vectors
$k_i$. As usual, the magnitude of 
$Z$ increases as $k_i$ is increased. 
This makes the problem rapidly convergent in our approach. 
The expansion coefficients for the current density in the basis $c\bN_0(k_ir)$ are shown
in Fig. (\ref{je}) (right panel) 
where the nature of convergence of our expansion is illustrated. 
The current density $j=j^c j_u$ is in units of $j_u=E_{ext}/\rho_u$  where the resistivity
unit is  $\rho_u= Z_0\omega at/c,$ $Z_0$ is the resistance of the vaccum. 
The IR experiments were carried out at a higher normalized wave vector and thus have more intermediate components. 
The magnitude of the current densities as a function of the normalized position on the cone for both the IR and the THz experiments  are shown in Fig. \ref{je} (left panel).
The charge density contains a bulk contribution from currents inside the cone 
given by $n=-i\nabla\cdot j/\omega$. 
Because the current density is nearly zero around the tip at $z/h\approx 0$, its spatial derivative and hence
this bulk charge density is only nonzero away from the tip at a distance much larger than $d$
since $d/h<<1$.
Thus this bulk charge will not contribute to near field results.
The total 
"bulk" charges  in units of $Q_u=\epsilon_0 E_{ext} a^2$ for the THz case 
[ (-0.07+0.04i) ]
and the IR
case [ (-0.16+0.007i) ] are comparable in magnitude. 


There are additional contributions to the charge from the surface charge densities  localized at the tip ($\sigma_{s2}$) and the base ($\sigma_{s1}$).
$\sigma_{si}=E_{si}\sigma_{u}$ where
$\sigma_{u}=2\epsilon_0E_{ext},$
For the IR vase,  $E_{s1}/E_{ext}=0.163-0.0168i,$ 
$E_{s2}/E_{ext}= 4.374+6.23i$. 
For the THz case
$E_{s1}/E_{ext}=0.142+0.188i,$ $E_{s2}/E_{ext}=-28.52-14.70i.$
The magnitude of the surface charge at the tip in the THZ case is five times bigger than that for the IR experiments. 
This comes about because the wavelength of the external field is much closer to
the size of the cone for the THZ experiments, $E_{i,ext}$ is dominated by the component with i=1, as is illustrated in Fig. \ref{je}.

\begin{figure}[tbph]
\vspace*{0pt} \centerline{\includegraphics[angle=0,width=9cm]{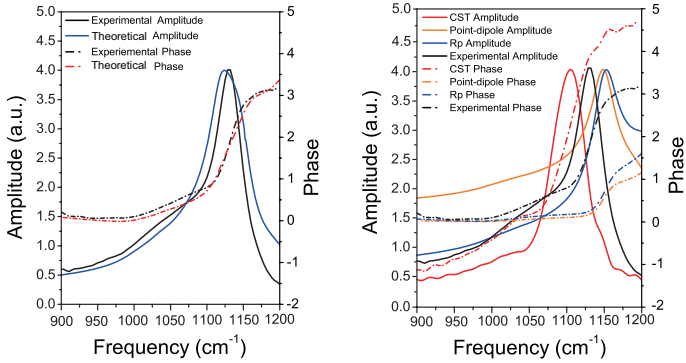}} 
\caption{ Left Panel:
Our theoretical and experimental 
results for the amplitude $S_2$ and the phase $\phi_2$ of uniform $SiO_2$ as a function 
of frequency. 
Right Panel: $S_2$ and $\phi_2$ calculated with the Rp model, the point-dipole model, and with the
CST numerical solver, together with the experimental result.}
\label{stne}
\end{figure}

We next look at the scattered field
$S_n$ at the modulated frequencies $\omega+n\Omega$ from the moving bulk and surface
charges. 
The scattered fields at a distance r from the tip
are proportional to the vector potential given by\cite{Jackson}
${\bf A}(r)=i\omega\mu_0e^{ikr} \int d{\bf r'}\bj(r')/r.$
The {\bf total} current density of the {\bf vibrating} tip is
a sum of a current induced by the incoming field and a current $j_d=n\partial_t d$
from the induced charge $n$ and the vibration of the tip $\partial_t d$.
$n$ contains both bulk and boundary charge contributions.
The "bulk" charges contribute mainly to the fundamental with n=1. 
The contributions to the higher harmonics with $n>1$ is dominated by near field
contributions when the tip is close to the surface and comes mainly from the localized charge at the tip. 
The radiation electric field $S_n$ is proportional to the 
Fourier transform at frequency $\omega+n\Omega$ 
of the contributions $j_{dl}$ to $j_d$ from the localized charges at the tip.
From Eq. (\ref{scd}) we get 
\beq
E \propto e^{i\omega t}(1+\beta) \cos(\Omega t)
/[1-2\beta I(d)].
\label{pt}
\eeq
As a test, we have calculated the amplitude $S_2$ and 
its phase as a function of frequency using as input dielectric constants
corresponding to a uniform surface of $SiO_2$.  The experimental result together with ours
are shown in the left panel Fig. (\ref{stne}).
The overall magnitude of the amplitudes 
and the baseline for the phase
depend on the experimental geometry 
and thus are the two adjustable parameters.
The agreement is very good. 
For comparison, results obtained
with commonly used simple tip modelling as well as CST simulation with a similar 
tip geometry 
are shown in the right panel.
The agreement with the experimental phase is not as accurate.

The fundamental is comparable 
and much larger than the higher harmonics respectively in recent experiments 
carried out at the THz and the IR frequencies.
The localized tip charge dominates the contribution to the higher harmonics and becomes 
much bigger for the THz experiments, 
providing an intuitive understanding of the origin of the near-field signals. 

We close with an example of the inverse problem of extracting the physical properties
from the measured\cite{expt} higher harmonics response for a gold disk of diameter approximately 
1.3 $\mu m$ and thickness $\approx 50 nm$.
We calculated a table of $S_3$ and $\phi_3$
as functions of $-0.1<Re[\beta]<1.4,$  $-0.5<Im[\beta]<0.5$ from eq. (\ref{pt})
for a mesh of 6400 grid points .
This takes 2.7 sec with a conventional Intel processor. 
For a pixel at position $\br$, we go through the entries 
in our table and identify $\beta(\br)$ as 
the one such that $|S_3(\beta(\br))-S_3^{expt}(\br)|+|\phi_3(\beta(\br))-\phi_3^{expt}(\br)|$
is minimized.
This calculation takes 0.77 s for a data set of $128\times 128$ pixels.

\begin{figure}[tbph]
\vspace*{0pt} \centerline{\includegraphics[angle=0,width=8cm]{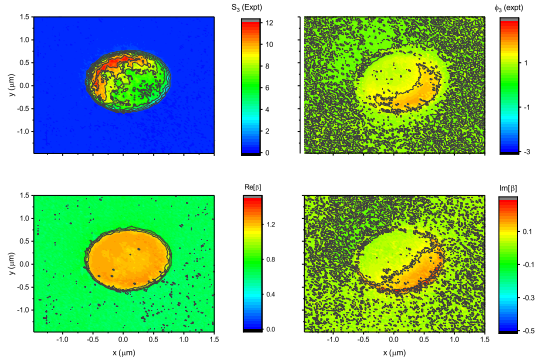}} 
\vspace*{0pt}
\caption{ Top Panels: The experimental amplitude and phase of the third harmonics  as a function of position on the disk.
Bottom Panels: The extracted values of $\beta$.}
\label{2d}
\end{figure}

Two dimensional plots of $\beta$, $S_3$ and $\phi_3$ are shown in Fig.  \ref{2d}.
Even though the material exhibits the same intrinsic $\epsilon$ independent of position,
there is a variation in the extracted $\beta$ from the average value.
We believe that is due to  charge densities on the disk induced by the EM field .
More precisely, $\beta$ is determined from the condition that
the change of the perpendicular component of the displacement field $D_{\perp}$ at the disk-air interface is equal to the surface charge denisty $\sigma_{disk}$:
$\Delta D_{\perp}=\sigma_{disk}.$ 
Our study of the EM scattering on disks\cite{disk} suggests that
there is an edge electric field and its associated charge at the 
perimeter of the disk.
The variation of the physical quantities is indeed dominated by changes at 
the perimeter, in agreement with our expectation.

\begin{figure}[tbph]
\vspace*{0pt} \centerline{\includegraphics[angle=0,width=8cm]{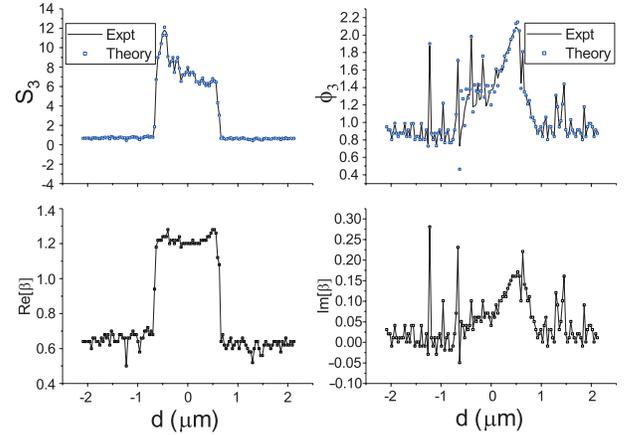}} 
\vspace*{0pt}
\caption{ Top Panels: The experimental amplitude and phase of the third harmonics 
together with the theoretical result for a metallic disk.
Bottom Panels: The extracted values of $\beta.$ These are as a function of position across a diameter of the disk.}
\label{disk13}
\end{figure}

To get a more quantitative picture of our results, we plot the physical quantities
as a function of the position across a diameter of the disk along 
the direction of the EM field
are shown in Fig. \ref{disk13}.
The experimental amplitude and phase of the third harmonics (top panels)
are well reproduced.
This substantiates the calculation reported in this paper.
The extracted $\beta$
are shown in the bottom panels.
We found that $Re[\beta]\approx 1.2$ (0), $Im[\beta]\approx 0.07$ (0) inside
(outside) the disk.  
This technique may also
be a new tool for mapping of the changes induced by external fields. 
The detailed analysis of this will be discussed in a separate paper.

In conclusion, we study the elastic scattering of an EM wave by a conical metallic surface via
representing the fields on the surface in basis functions containing the integrable
singularities at the tip. We apply our calculation to s-SNOM and found good agreement
with experimental results from a vibrating metallic tip over a uniform sample. We found
the field-induced charge distribution consists of a bulk term along the body of the cone
and localized terms at the tip apex and the base. The bulk charge is far away from the tip
and contributes little to the near-ﬁeld higher harmonics in the scattered field. The localized boundary charges at the tip apex contribute to the near-field signal and correspond to the charge accumulation effects previously discussed in pn junctions and on the surface of capacitor plates in AC experiments\cite{kittel,ac,hebard}. In these treatments, the additional physics of the diffusion of the electrons are included and a finite width in the localized charge distribution of the order of the screening length is found. This localized boundary charge provides new insight into the physics of the scattering problem, which can be related to the distinct $S_1/S_2$ 
ratio found in recent experiments at IR and THz frequencies. In addition, we also demonstrate the application of our technique by extracting a two-dimensional effective dielectric constant map of a finite plasmonic metallic disk from experimental data, which is non-uniform due to the induced charge density distribution under the illumination of EM field. Our calculations of the scattering signal and the back-extraction process are
more realistic and much faster compared to the prevailing analytical and numerical approaches.
Future studies will include the calculation of near-field scattering signal of nonuniform samples. 


\begin{thebibliography}{99}
\bibitem{cst}
Numerical calculations with the commerical software "COMSOL"  was
reported by many authors but we could not find any published result for the phase.
We have carried out calculations with a comparable commerical oftware
"CST Studio" and the amplitude and the phase is reported below.
\end{thebibliography}
\end{document}